\documentclass[]{spie}  

\usepackage{amsmath,amsfonts,amssymb}
\usepackage{graphicx}
\usepackage[colorlinks=true, allcolors=blue]{hyperref}
\usepackage{astro_bib_macro}

\title{James Webb Space Telescope Optical Simulation Testbed V: Wide-field phase retrieval assessment}

\author{Iva Laginja\supit{a}, Greg Brady\supit{a}, R\'{e}mi Soummer\supit{a}, Sylvain Egron\supit{g}, Christopher Moriarty\supit{a}, Charles-Philippe Lajoie\supit{a}, Aur\'{e}lie Bonnefois\supit{b}, Vincent Michau\supit{b}, \'{E}lodie Choquet\supit{e}\supit{f}, Marc Ferrari\supit{c}, Lucie Leboulleux\supit{a}\supit{b}\supit{c}, Olivier Levecq\supit{h}, Mamadou N'Diaye\supit{d}, Marshall D. Perrin\supit{a}, Peter Petrone\supit{a}, Laurent Pueyo\supit{a}, Anand Sivaramakrishnan\supit{a}
\skiplinehalf
\supit{a} Space Telescope Science Institute, 3700 San Martin Drive, Baltimore, MD 21218, USA\\
\supit{b} Office National d'\'{E}tudes et de Recherches A\'{e}rospatiales, 29 Avenue de la Division Leclerc, 92320 Ch\^{a}tillon, France\\
\supit{c} Aix Marseille Universit\'{e}, CNRS, LAM (Laboratoire d'Astrophysique de Marseille) UMR 7326, 13388 Marseille, France\\
\supit{d} Observatoire de Nice C\^{o}te d'Azur, Boulevard de l'Observatoire, 06304 Nice, France\\
\supit{e} Department of Astronomy, California Institute of Technology, 1200 East California Boulevard, MC 249-17, Pasadena, CA 91125, USA\\
\supit{f} Jet Propulsion Laboratory, California Institute of Technology, 4800 Oak Grove Drive, Pasadena, CA 91109, USA\\
\supit{g} Iridescence S.A.R.L., 149 avenue du Maine, 75014 Paris, France\\
\supit{h} DAMAE medical, 28 Rue de Turbigo, 75003 Paris, France
}

\authorinfo{Further author information, send correspondence to Iva Laginja: E-mail: ilaginja@stsci.edu, Telephone: 1 667 218 6530}

\begin{document} 
\maketitle

\begin{abstract}
The James Webb Space Telescope (JWST) Optical Simulation Testbed (JOST) is a hardware simulator for wavefront sensing and control designed to produce JWST-like images. A model of the JWST three mirror anastigmat is realized with three lenses in the form of a Cooke triplet, which provides JWST-like optical quality over a field equivalent to a NIRCam module. An Iris AO hexagonally segmented mirror stands in for the JWST primary. This setup successfully produces images extremely similar to expected JWST in-flight point spread functions (PSFs), and NIRCam images from cryotesting, in terms of the PSF morphology and sampling relative to the diffraction limit. The segmentation of the primary mirror into subapertures introduces complexity into wavefront sensing and control (WFS\&C) of large space based telescopes like JWST. JOST provides a platform for independent analysis of WFS\&C scenarios for both commissioning and maintenance activities on such observatories. We present an update of the current status of the testbed including both single field and wide-field alignment results. We assess the optical quality of JOST over a wide field of view to inform the future implementation of different wavefront sensing algorithms including the currently implemented Linearized Algorithm for Phase Diversity (LAPD). JOST complements other work at the Makidon Laboratory at the Space Telescope Science Institute, including the High-contrast imager for Complex Aperture Telescopes (HiCAT) testbed, that investigates coronagraphy for segmented aperture telescopes. Beyond JWST we intend to use JOST for WFS\&C studies for future large segmented space telescopes such as LUVOIR.\\
\end{abstract}

\keywords{Segmented telescope, cophasing, exoplanet, high-contrast imaging, error budget, wavefront sensing and control}

\section{INTRODUCTION}
\label{sec:INTRODUCTION}

For large segmented telescopes, there is the need to actively control the telescope in order to achieve the optimum alignment and optical quality, bringing the primary from millimeter misalignments to fine alignments of nanometers. While controlled optics have become common on ground based telescopes, this technique will be extended to space for the James Webb Space Telescope (JWST). While its predecessor, the Hubble Space Telescope (HST), consisted of a mostly passive design with the exception of a variable defocus of the primary mirror, JWST will have 132 degrees of freedom between the primary and the secondary mirror. These will be initially aligned during the commissioning activities of the telescope, then maintained by periodic wavefront sensing and control (WFS\&C) activities during its lifetime of at least five to ten years in order to maintain superb image quality\cite{acton2004, acton2012wfsc-for-jwst, knight2012align}. The procedures for WFS\&C on the JWST have been thoroughly tested in simulation and experiment \cite{barto2008} on the Testbed Telescope (TBT) at Ball Aerospace \cite{acton2006, acton2007}, a 1:6 scale model of the telescope that is equipped with the same degrees of freedom as the original, as well as the Integrated Telescope Model (ITM) software \cite{knight2012itm}. Most recently, WFS\&C methods have been demonstrated on the integrated flight hardware \cite{acton2018, lajoie2018}. However, during the decade-plus development of JWST, WFS\&C algorithms have continued to develop, and new advanced algorithms are worth investigating to expand the toolkit for alignment and maintenance of JWST.

The JWST Optical Simulation Testbed (JOST) at the Space Telescope Science Institute (STScI) provides a platform to test such algorithms for segmented mirror control, evaluating them for possible applications on JWST as well as on future space missions with active primary segmentation like the Large UV/Optical/IR Surveyor (LUVOIR) \cite{dalcanton2015, bolcar2017}. It is a simplified tabletop model of the JWST, as opposed to a high-fidelity scaled model like the TBT, but it is a close enough physical representation to model the key optical aspects. It is a supplement to existing verification and validation activities for independent cross-checks and novel experiments, not a part of the mission’s critical path development process. In addition to exploring phase retrieval methods and implementing linear wavefront control over a wide field of view, JOST is used to develop staff expertise for commissioning and operations, conveniently being co-located at STScI with the Science \& Operations Center (S\&OC) that will support commissioning and be responsible for operations of the JWST.

 JOST is a three lens anastigmat, a refractive analogue to JWST's three mirror anastigmat. An aperture stop defines the system's pupil while the segmentation is provided by the planar segmented deformable mirror, whose segments can be controlled in piston, tip and tilt. The secondary lens (L2) that stands in as surrogate for JWST's secondary mirror is  motorized in tip and tilt, and x, y and z translation. JOST has in total 59 motorized degrees of freedom, which are the most relevant ones for WFS\&C maintenance activities. The setup design meets the requirement of an image quality of a minimum wavefront error of 40 nm rms at a wavelength of 638 nm over a field equivalent to one NIRCam module, and our latest measurements confirm that we meet this requirement, as we detect a minimum wavefront error rms of 15 nm.

This paper presents the optical characterization of JOST's full field of view after the successful fine alignment of L2 and DM, done previously. Our group presented a general overview of JOST in Perrin et al. \cite{perrin2014} (hereafter paper I). Its detailed optical design and several trade studies were presented in Choquet et al. \cite{choquet2014} (hereafter paper II). The experimental implementation of the WFS\&C on the testbed is described in Egron et al. 2016 \cite{egron2016} (hereafter paper III). The experimental results regarding the linear control of L2 are described in ICSO 2016 proceeding Egron et al. \cite{egron2016icso} (hereafter Egron ICSO) and the alignment of the segmented deformable mirror is presented in Egron et al. 2017 \cite{egron2017} (hereafter paper IV). Before moving on to the implementation of WFS\&C algorithms beyond the linearized algorithm for phase diversity (LAPD), we perform wide-field wavefront sensing on JOST. In section \ref{sec:TESTBED DESCRIPTION} we give an overview of the testbed and its recent changes in hardware and software. In section \ref{sec:WIDE-FIELD WFS WITH DM} we describe the wide-field wavefront sensing and present its results, and finally we summarize and conclude our findings in section \ref{sec:CONCLUSION}.

\section{TESTBED DESCRIPTION}
\label{sec:TESTBED DESCRIPTION}

\begin{figure}
   \begin{center}
   \begin{tabular}{c}
   \includegraphics[height=8cm]{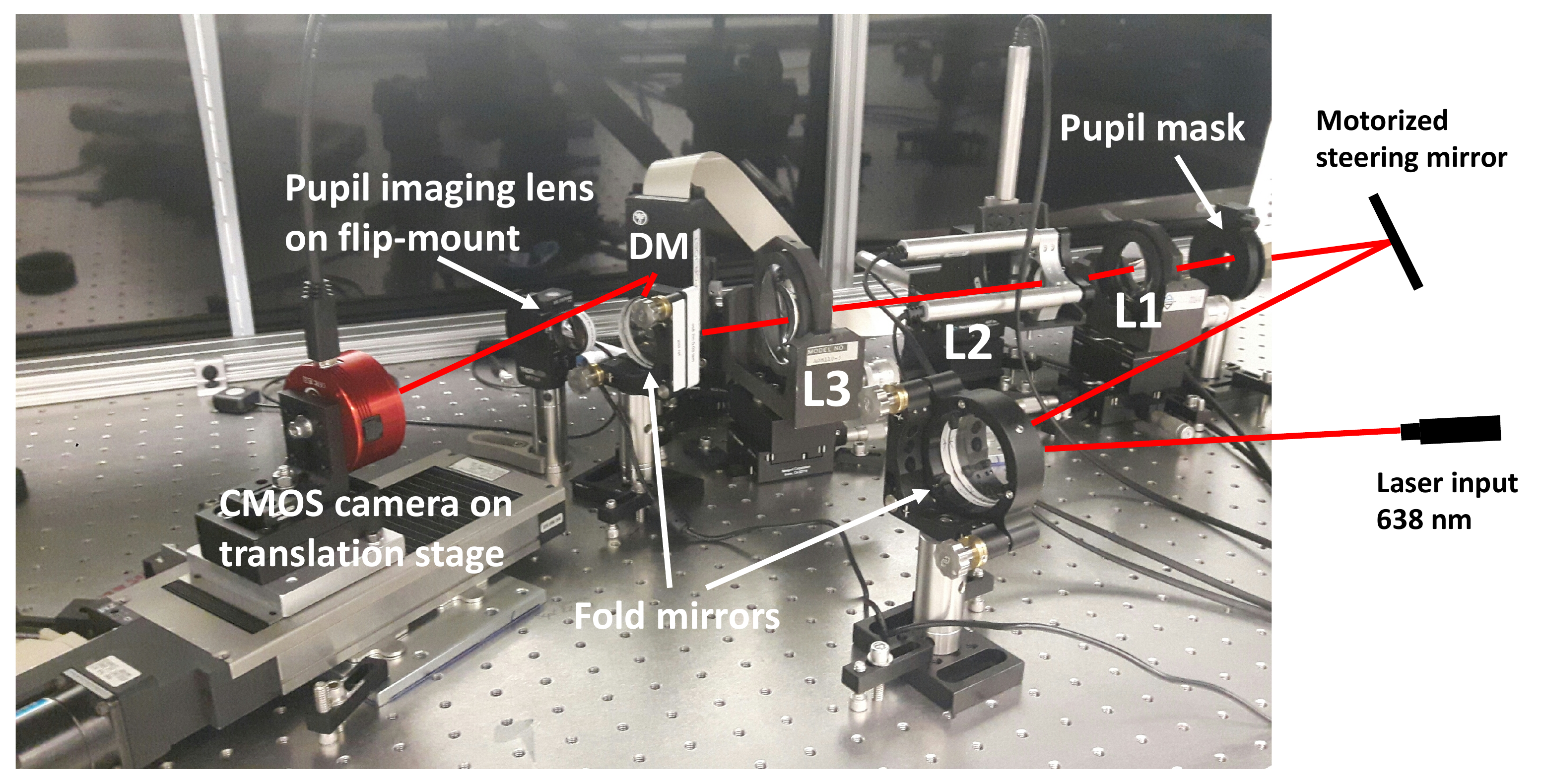}
   \end{tabular}
   \end{center}
   \caption[Testbed layout] 
   { \label{fig:JOST_setup} 
JOST testbed layout. An off-axis parabola (OAP, not pictured) and a fold mirror put the laser beam launched from an optical fiber onto a steering mirror that is controlled in x and y by a stepper motor. This mirror illuminates the JWST-like pupil mask which directs the beam into the telescope simulator. The three lenses L1, L2 and L3 form a Cooke Triplet, a refractive analogue to the reflective three-mirror anastigmat (TMA) of the JWST. L2 acts as a surrogate for the JWST secondary mirror and is independently controllable by motors in tip and tilt, and x, y and z translation. Another fold mirror positions the beam on a subset of 18 segments of the Iris AO segmented deformable mirror, where all the segments can be independently controlled in piston, tip and tilt. A pupil imaging lens is attached to a flip-mount which allows for a fast change between pupil and focal plane imaging mode. The camera is mounted on a translation stage of 100 mm travel, which enables us to take phase-diverse data sets. The testbed can accommodate either a CMOS (shown here) for faster acquisition and smaller field of view, and a CCD providing the same field of view as a a NIRCam module, and identical sampling.}
   \end{figure}

An extensive description of the JOST optical design can be found in paper I \cite{perrin2014} and paper II \cite{choquet2014}.  The updated current layout can be seen in Fig. \ref{fig:JOST_setup}. The main components of JOST are a fiber launch, steering mirror for wide-field exploration, a JWST-like pupil mask, a telescope simulator made of three custom lenses, a segmented deformable mirror (DM) and a camera on a translation stage, to be able to provide focus-diverse images.

\subsection{Key hardware components}

\begin{figure}
   \begin{center}
   \begin{tabular}{c}
   \includegraphics[height=11cm]{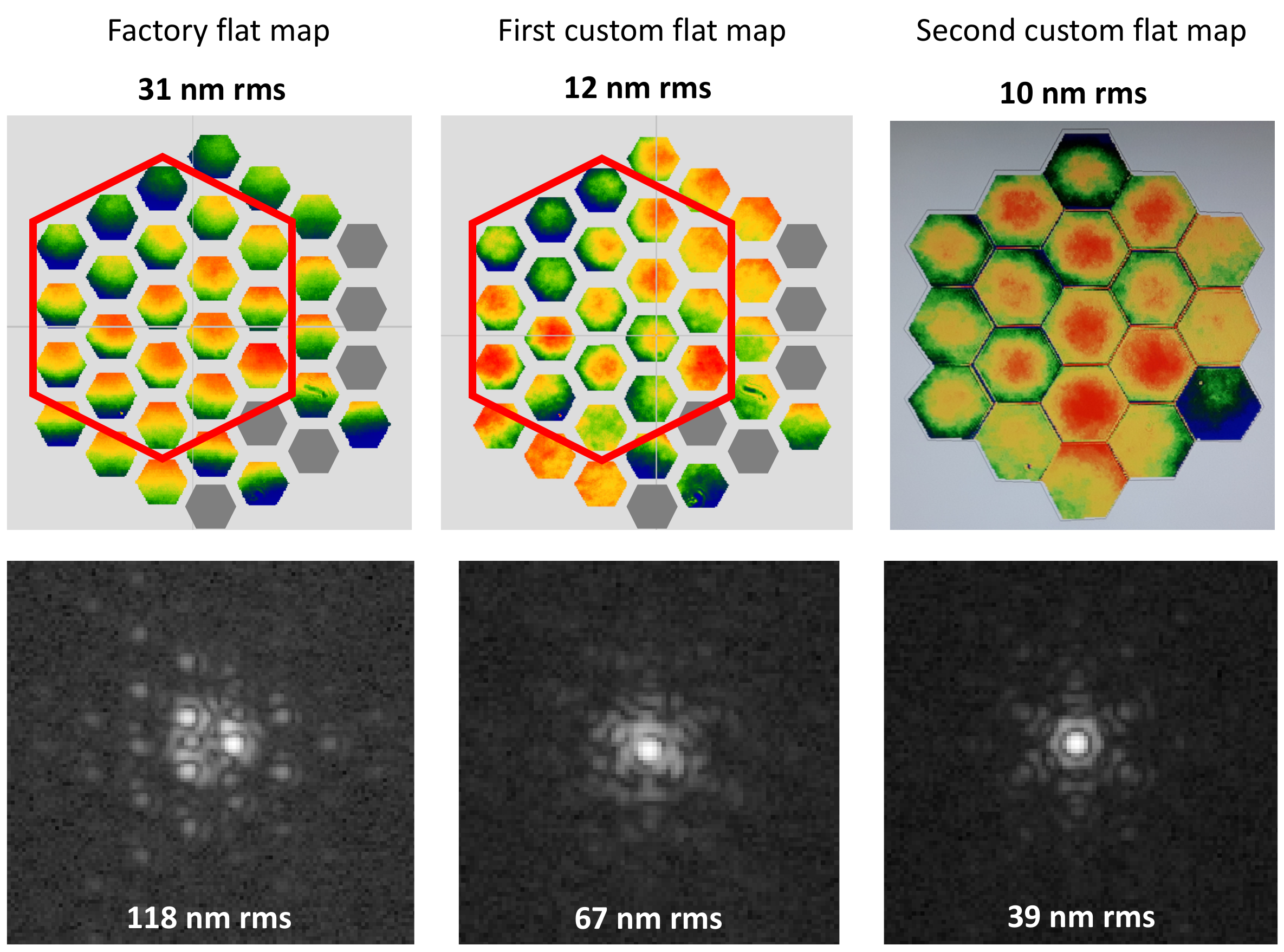}
   \end{tabular}
   \end{center}
   \caption[New flat map] 
   { \label{fig:DM_calibration} 
Comparison of different flat maps for the segmented DM. \textit{Top row:} Interferometer measurements of the DM surface with different flat map implementations in the Iris AO GUI and the resulting surface flatness rms. The red hexagon in the two left pictures denotes the JOST subaperture, while the third picture shows the JOST subaperture only. Grey hexagons indicate the position of dead segments in the engineering-grade device used on JOST. \textit{Bottom row:} The corresponding PSFs to each flat map and their wavefront error rms. \textit{Left column:} Vendor-provided flat map where all values for piston, tip and tilt in the mirror GUI are zero. The DM was initially calibrated in the horizontal plane, lying down flat, while we use it standing upright in a mount, which makes all the segments sag forward and introduce a large local tilt. This leads to a PSF that shows many individial sub-PSFs, as the segments are not stacked properly. \textit{Middle:} First custom flat map from 2016. The segments have been calibrated with masks that ignore the interfaces between the individual segments and only with global Zernike mode adjustments. The PSF is stacked, but still shows residual aberrations in the overall PSF structure. \textit{Right:} Second and current custom flat map from late 2017. Instead of global modes, individual adjustments of piston, tip and tilt have been made on each segment. Further, the mask to define the JOST pupil also includes the discontinuities in the segment gaps, giving a better estimate of the final performance.}
   \end{figure}

\begin{figure}
   \begin{center}
   \begin{tabular}{c}
   \includegraphics[height=6.3cm]{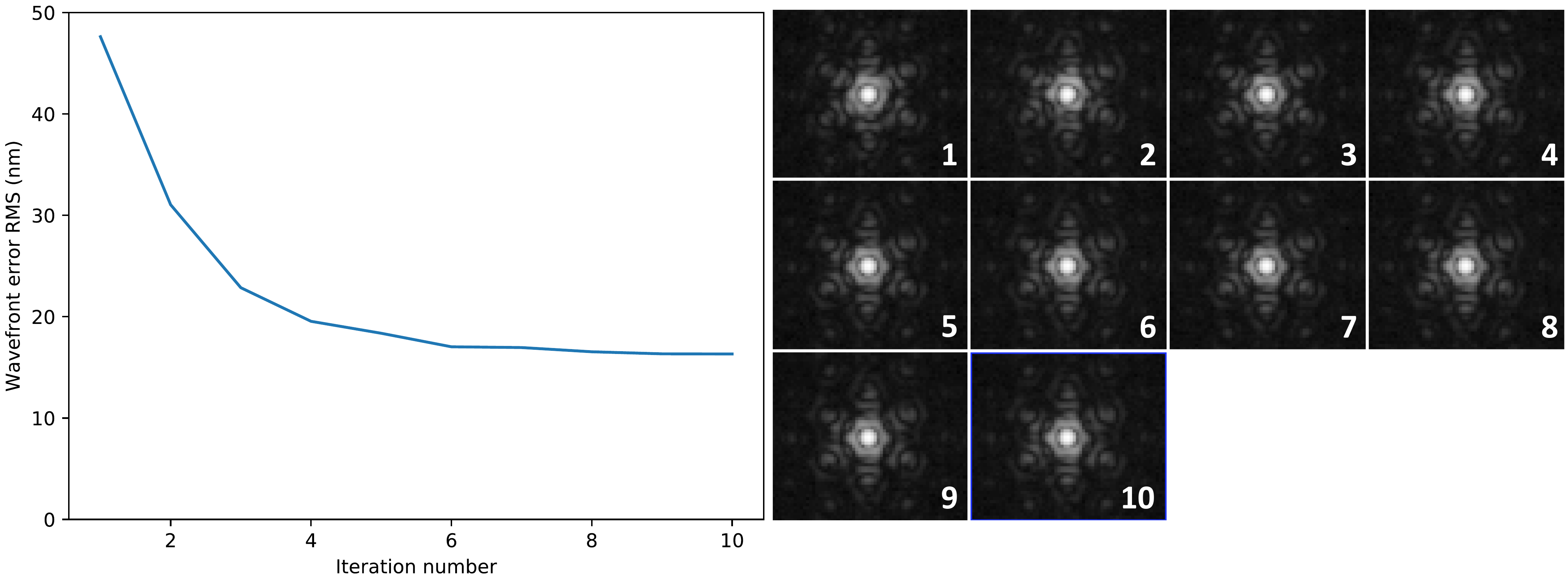}
   \end{tabular}
   \end{center}
   \caption[LAPD iterations] 
   { \label{fig:LAPD_iterations} 
Closed loop LAPD WFS\&C experiment with the DM segment controls as control modes. \textit{Left:} rms wavefront error vs. iteration number, \textit{right:} PSF at each iteration step. Iteration 1 is the initial configuration of the testbed, with each consecutive iteration of LAPD wavefront sensing and linear control of the DM providing a better overall quality of the system. The rms wavefront error drops to 16 nm after only six iterations.}
   \end{figure}

The segmentation of the testbed, including gaps of the same size ratio like on JWST, is provided by an Iris-AO segmented deformable mirror. The entire mirror has 37 independently controllable segments in piston, tip and tilt. A conjugated pupil mask with a hexagonal central obscuration and spiders defines the area of 18 segments that ultimately form the pupil. The DM can be controlled either by a GUI that is provided by the company, or by directly using the application programming interface (API) written in the C programming language, which we can call from within a Python wrapper. For best performance of the testbed, we need to have a baseline flat map of the DM, which is the configuration of the segments that gives the best optical  quality of our data. The DM calibration provided by the vendor was for the DM oriented in the horizontal plane, facing upwards, while we use it standing upright in a mount facing horizontally, which makes all the segments sag forward and introduce large local tilts. A first flat map was created by using the GUI and checking the results directly with a Fizeau interferometer, in 2016. While this flat map showed a major improvement over the factory-defined flat position of the DM, we were able to do an even more accurate calibration in late 2017 by adjusting each segment individually instead of using only global modes. The improvements led to an overall surface flatness of 10 nm rms over the entire pupil of 18 segments, creating a PSF with 39 nm rms wavefront error. The calibration maps and the PSFs we observe when the according flat map is put on the DM can be seen in Fig. \ref{fig:DM_calibration}. This flat configuration can be improved further by running closed-loop iterations of LAPD on it, with the segment piston, tip and tilt being the controlled modes of this WFS\&C experiment. Fig. \ref{fig:LAPD_iterations} shows how the overall wavefront error rms value drops from 40 nm to 16 nm after six iterations.

The telescope simulator consists of three custom made lenses that form a Cooke triplet, a refractive analogue to the JWST three mirror anastigmat, providing good optical quality over a wide field of view. During the alignment process of JOST, the third lens (L3) was added to the setup in reverse \cite{egron2016}. The main effect of this is to reduce the size of the well-corrected field of view, but since it does not change the general alignment physics, and for convenience of use (fast readout, shorter exposure times, low noise) we have switched to a CMOS camera, whose smaller field of view matches the performance of the current system well, we have decided to keep this configuration for now, until all wide-field infrastructure is finalized. Both cameras can be readily interchanged. 


In order to be able to provide focus-diverse images which are needed for phase retrieval, the camera is mounted on a translation stage which provides a movement range of 100 mm in the z-direction (optical axis). The initial setup of JOST was using a CCD camera. However, the closed-loop experiments for wavefront sensing algorithms require 20 to 100 images be taken during each loop, and with the exposure times of one to ten seconds of the CCD camera the readout overheads would take a disproportionally long time compared to the wavefront sensing computations. We replaced the CCD with a ZWO monochrome CMOS camera that operates with exposure times between 0.5 and 500 ms, thus making the process of image acquisition considerably faster. The pixel size of the new camera is almost a third of the old size, 3.7 microns versus 9 microns, but a 2x2 pixel binning in the data reduction process leaves us with an effective pixel size of 7.4 microns, which means that our focused PSFs are still sampled with a factor slightly over 2, albeit not at the exact same sampling as NIRCam, but this has no impact on algorithm development, and both camera setups remain available. 

The before-mentioned piston, tip and tilt controls of the segmented DM over all 18 segments total 54 degrees of freedom on the DM. The secondary surrogate lens L2 can be remotely controlled with a stepper motor in tip and tilt, and x, y and z translation, yielding a total of 59 motorized degrees of freedom on JOST that are used in the linear control model \cite{egron2016}. While JWST has a total of 132 degrees of freedom, the ones JOST misses are clocking, radius of curvature adjustment, and x and y translation of each individual segment, all of which are control modes whose major adjustments will happen during the initial six month commissioning process after launch. The degrees of freedom which are the same between JOST and JWST on the other hand are those with the largest optical influence functions during wavefront maintenance activities.

There is a steering mirror, movable in tip and tilt, positioned right before the JOST pupil mask. With this motorized mirror we have the possibility of directing the laser beam to off-axis positions on the detector and exploring field-dependent aberrations.

\subsection{Software upgrades}

Between October 2017 and April 2018, effort was put into restructuring the JOST software components. The first and crucial step was to migrate all prototype software to a version control system, to allow for a smoother and safer way of collaboration. Furthermore, all code involved in JOST data acquisition, data reduction and hardware control have been translated to and extended in Python\footnote{\url{https://www.python.org}}.

JOST is co-located with the High-contrast imager for Complex Aperture Telescopes \cite{soummer2018} (HiCAT) and the two testbeds are taking advantage of each other's developments. The HiCAT experiment control software was recently rearchitectured as a modern, clean Python package by a professional software engineer. Each instrument and hardware component has an interface that is easily accessible by simply installing the HiCAT package and importing it. Each hardware interface follows a simple object-oriented paradigm where the parent is an abstract class (e.g. "Camera"), which defines specific methods and implements a context manager. Context managers are important for hardware control because they will gracefully close the hardware even if the program crashes unexpectedly. The child classes implement the abstract methods such as open(), close(), takeExposure() with code for the specific camera. This keeps the scripts generic and means changing cameras will have little to no impact on the code. A thorough description of the HiCAT software infrastructure is given by Moriarty in these proceedings \cite{moriarty2018}.

We installed and started using the HiCAT package on JOST. Since JOST uses the same type of hardware like HiCAT (same laser source and motor controllers, same camera type but different model), we only needed to update the configuration file of the package and were able to use the code as is. The modular structure of the code allows for very fast and clean generation of new scripts and implementation of new experiments. These changes are intended to push for best practices in astronomy coding, and incidentally move away from more traditional programming languages used in astronomy like IDL and Mathematica, providing a concise environment for the work done. This arrangement will facilitate JOST's role in providing a flexible multipurpose laboratory testbed for the testing and validation of independent phase retrieval techniques.

\subsection{Previous wavefront sensing and control activities}

In paper III, the authors describe the alignment of the three lenses of JOST with a phase diversity algorithm provided by the Office National d'Etudes et de Recherches Aerospatiales (ONERA) in France, and a linear optical control model. At that point, JOST did not yet include the segmented deformable mirror. Their results show a symmetric degradation of the wavefront when moving away from the optical axis. In the following paper IV \cite{egron2017}, a linearized algorithm for phase diversity (LAPD) \cite{mocoeur2009} was used for the cophasing of the newly inserted segmented DM with the aligned testbed. In this algorithm, the pupil is made out of 18 hexagonal subapertures that simulate the effective JOST pupil consisting of the pupil mask and the DM segmentation, while the previous algorithm used for the lens alignment was working with a circular pupil without any obscuration. LAPD allowed for the alignment of the 18 mirror segments in piston, tip and tilt on each segment individually, and having both the mirror and the lenses aligned left the total wavefront error of JOST with an rms of under 40 nm.

Paper IV completed the full automation of the JOST testbed with regards to hardware control, data acquisition and reduction, wavefront sensing with an arbitrary phase retrieval algorithm and wavefront control with a linear coupling model. While the WFS\&C in paper III was implemented on a wide field of view, covering a range of (1$^\circ$, -1$^\circ$), the WFS\&C after the addition of the segmented mirror was performed only on-axis.

\section{WIDE-FIELD WAVEFRONT SENSING WITH A SEGMENTED DEFORMABLE MIRROR}
\label{sec:WIDE-FIELD WFS WITH DM}

\subsection{Goals of wide-field WFS demonstration}
\label{subsec:goals_wide-field}

The new goal is to expand the testbed capabilities to operation on a wide field for all degrees of freedom of JOST, which means we want to implement a wide-field approach to WFS\&C on both the degrees of freedom of the DM (18 segments times 3 modes) and the motorized L2 variations (x, y and z translation plus tip and tilt). To achieve this we are seeking a validation of the L2 alignment with a hexagonally segmented pupil (as opposed to the round pupil in paper III) and a closed loop WFS\&C performance on an extended field of view with the DM. In this paper we present the results of wide-field wavefront sensing with LAPD and a characterization of the JOST field of view now that the DM is in place.

Since our camera changed from a CCD to a CMOS camera, our field has shrunk to 0.6$^\circ$ x 0.9$^\circ$, but we extended our characterization beyond that by translating the camera radially away from the optical axis. With the motorized steering mirror in place it is easy to find the optical axis at any time and it enables us to iteratively explore a wide field of view.

\subsection{Data acquisition and reduction}
\label{subsec:data-acquisition-reduction}

To be able to run any kind of phase retrieval, we need to make an in-focus image and at least one defocused image of the testbed configuration we are interested in. On JOST, the defocus diverse data is acquired by moving the camera on a translation stage; we move the camera by 93 mm to introduce a defocus of 4 waves (23.88 rad). The nominal field of view of the camera is 0.6$^\circ$ x 0.9$^\circ$, so in order to get images out to 1.0$^\circ$, we move the camera to different lateral offsets. A one-time shift is not enough though: the further away we move from the optical axis, the bigger the offset on the camera between the focused and defocused images will be, so the range of field points we can cover with one camera offset gets smaller with every step further outward. With five offsets, we cover a lateral distance from -0.3$^\circ$ to 1.02$^\circ$ and at one offset position, we move the steering mirror to 20 different distances from the optical axis and obtain 20 focused and defocused images each, as well as background images for both camera positions, thus probing the field in 100 different positions.

An automated Python script processes the images through the standard steps of stacking, background subtraction, centering, 2x2 binning (leaving us with 512 x 512 pixel images), bad pixel correction and normalization. The result are one focused and one defocused image that are consequently used by the wavefront sensing code to determine the wavefront aberrations at all points in the field of view. We obtain the overall wavefront error rms values through LAPD, which also creates wavefront maps which we then decompose with the Python package POPPY \cite{perrin2012} into individual Zernike modes from defocus (Z4; we follow the Noll convention \cite{noll1976} for Zernike numbering) to primary spherical (Z11).

\subsection{Wavefront sensing results}
\label{subsec:results}

\begin{figure}
   \begin{center}
   \begin{tabular}{c}
   \includegraphics[height=7cm]{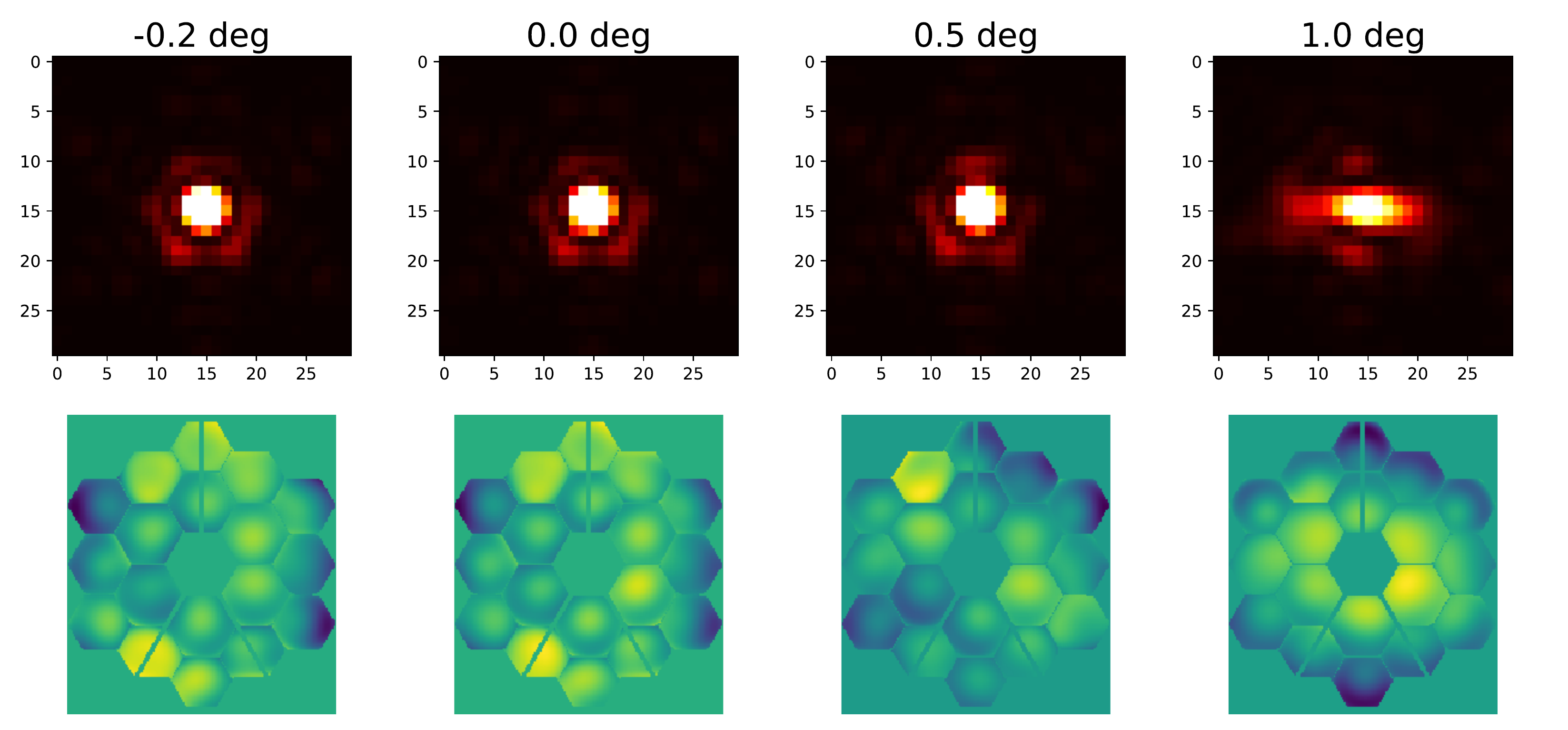}
   \end{tabular}
   \end{center}
   \caption[Wavefront maps and PSFs] 
   { \label{fig:wavefront-map} 
 Point spread functions (PSFs) and wavefront maps of four different points in lateral translation from the optical axis. The wavefront quality is very uniform out to a distance of 0.5$^\circ$, the PSFs look very much alike. While the wavefront maps of the PSF at -0.2$^\circ$ and 0.0$^\circ$ look very similar, the map of the PSF at 0.5$^\circ$ is starting to show global aberrations, as inferred by the darker areas in the top right and bottom left parts of the pupil. The PSF at 1.0$^\circ$ away from the optical axis is very clearly aberrated, showing very strong astigmatism, which is also confirmed in the Zernike decomposition of the wavefront maps in Fig. \ref{fig:contour-total}. The according wavefront map also distinctly shows astigmatism and defocus.}
   \end{figure}

In a first step, we inspected the point-spread functions (PSFs) and wavefront maps obtained by the data acquisition and LAPD wavefront sensing. In order to clearly see an aberrated PSF, one has to go to a field point well beyond 0.5$^\circ$. The rightmost PSF frame in Fig. \ref{fig:wavefront-map} shows the distorted PSF at 1.0$^\circ$, and in this image, astigmatism is very clearly seen. The wavefront maps show the wide-field aberrations a bit earlier, e.g. at 0.5$^\circ$. The center right wavefront image in Fig. \ref{fig:wavefront-map}, at a distance of 0.5$^\circ$, is starting to show a global tendency of the dark and bright wedges typical for astigmatism. This global wavefront patterns becomes more distinct when looking at the far right wavefront map in Fig. \ref{fig:wavefront-map}, at a field point of 1.0$^\circ$.

\begin{figure}
   \begin{center}
   \begin{tabular}{c}
   \includegraphics[height=10cm]{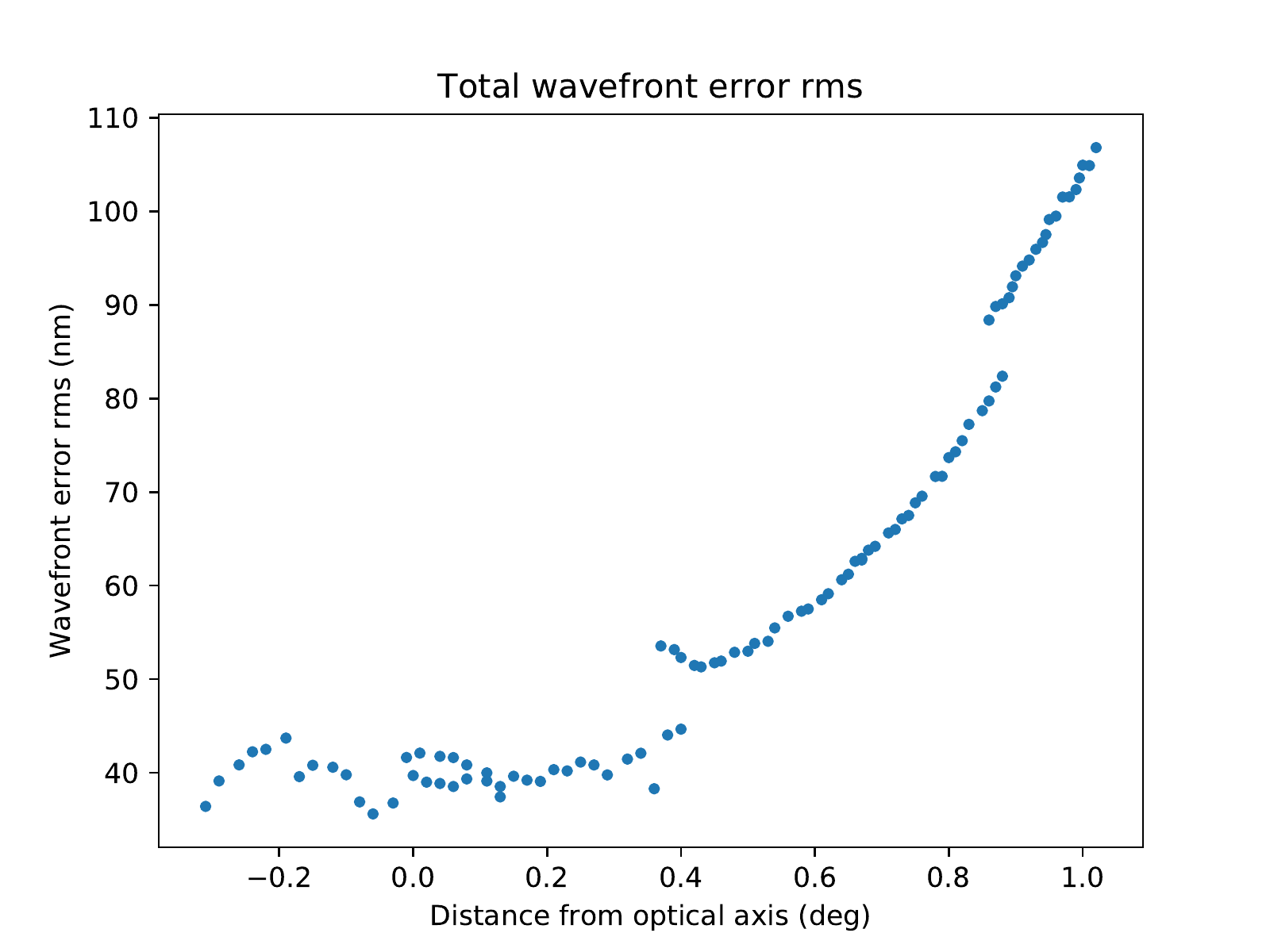}
   \end{tabular}
   \end{center}
   \caption[Total rms radial] 
   { \label{fig:contour-total} 
 Total wavefront error rms of radially translated PSFs on JOST, as retrieved by LAPD, taking into account the shape and segmentation of the JOST pupil. The wavefront error is relatively uniformly scattered around 40 nm rms all the way out to 0.4$^\circ$, with a significant rise after 0.5$^\circ$, from where on it continuously grows to over 100 nm rms at 1.0$^\circ$ away from the optical axis. The discontinuities in the data at 0.4$^\circ$ and 0.85$^\circ$ are the limits between two data sets that are separated by a shift of the camera, as described in Sec. \ref{subsec:data-acquisition-reduction}. Since the optical properties of the testbed will be radially symmetric around the optical axis, this shows that JOST has very good optical quality in a wide field of 1$^\circ$ x 1$^\circ$, centered on the optical axis of the system. Note that our baseline wavefront error rms here is 40 nm, while we demonstrated in Fig. \ref{fig:LAPD_iterations} that we can align the testbed to about 15 nm rms. While in Fig. \ref{fig:LAPD_iterations} we demonstrate that we have the ability to go down to 15 nm, we did not change the baseline alignment of the testbed to match our best alignment  state of 15 nm rms in this current wide-field characterization.}
   \end{figure}

In Fig. \ref{fig:contour-total}, we can see how the overall wavefront error changes as a function of distance from the optical axis of the testbed. The wavefront error is relatively uniformly scattered around 40 nm rms until a radial distance of 0.4$^\circ$ and it starts to rise significantly beyond 0.5$^\circ$. This confirms that the JOST anastigmat has good optical quality in a field of view with a diameter of 1$^\circ$. There are some discontinuities appearing in the data around 0.4$^\circ$, 0.66$^\circ$ and 0.85$^\circ$, which indicate the interface between two datasets that have been taken before and after a lateral detector shift as described in Sec. \ref{subsec:data-acquisition-reduction}.

\begin{figure}
   \begin{center}
   \begin{tabular}{c}
   \includegraphics[height=10cm]{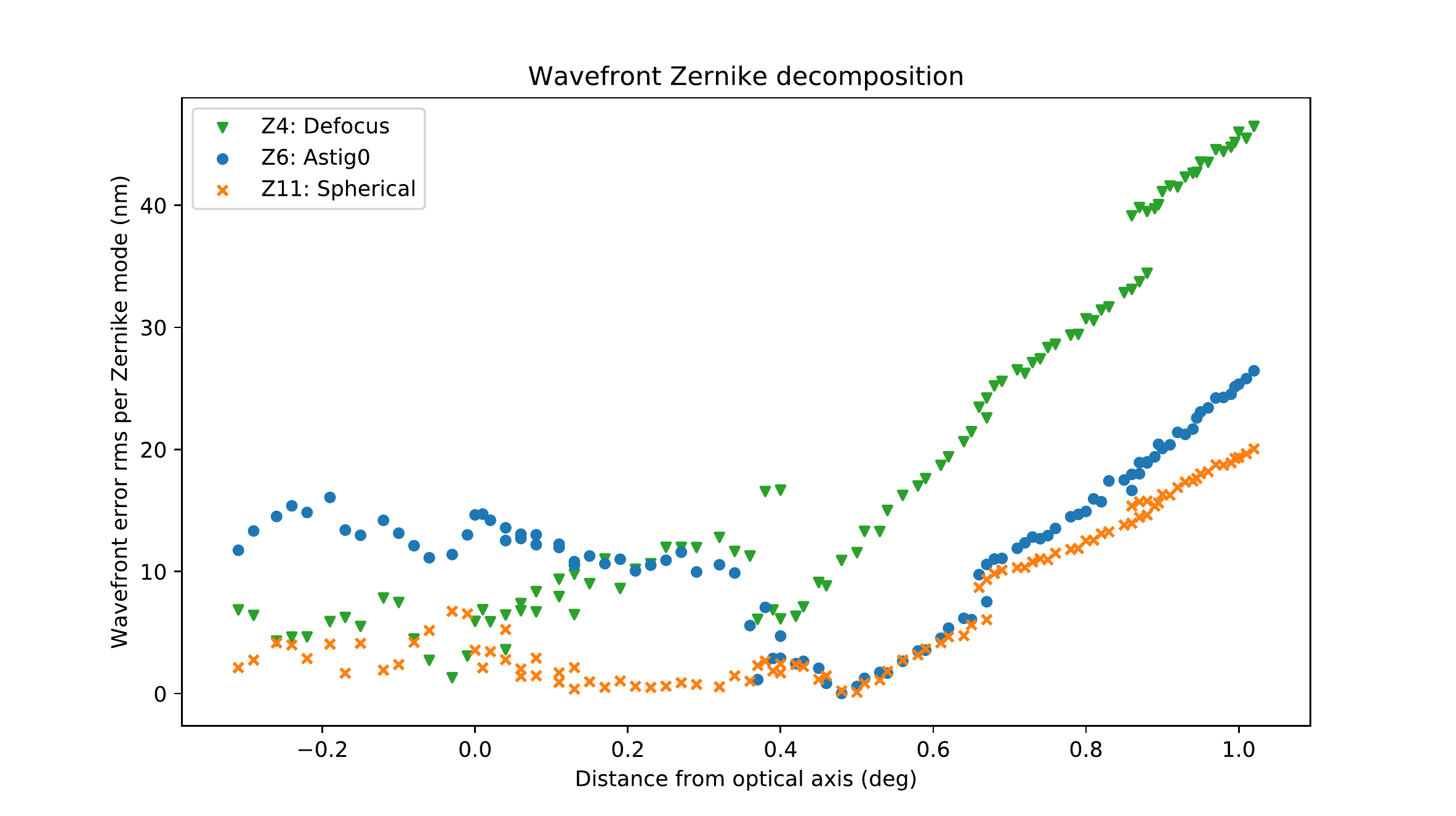}
   \end{tabular}
   \end{center}
   \caption[Zernike decomposition] 
   { \label{fig:zernikes} 
 Decomposition of the wavefront error of radially translated PSFs on JOST into individual Zernike modes. We probed the wavefront from piston (Z1) to spherical aberration (Z11) and are showing here the three most influential modes, defocus, 0$^\circ$ astigmatism and spherical aberration. The discontinuities in the data at 0.4$^\circ$, 0.65$^\circ$ and 0.85$^\circ$ are the limits between two data sets that are separated by a shift of the camera, as described in Sec. \ref{subsec:data-acquisition-reduction}. As is the total wavefront error rms (see Fig. \ref{fig:contour-total}), the Zernikes remain flat throughout the PSFs until 0.4$^\circ$ off the optical axis. Beyond that, their contribution to the total wavefront error grow continuously throughout all the data. From paper II and III we would expect a lot of defocus, astigmatism and coma, but not as much spherical aberration. We are currently investigating where this is coming from.}
   \end{figure}

Using the wavefront maps generated by LAPD, we decomposed each individual wavefront at each field point into the 11 first Zernike modes of the Noll convention. The three modes contributing the strongest to the overall wavefront error have been found to be Z4, Z6 and Z11 - defocus, 0$^\circ$ astigmatism and 3$^{rd}$ order spherical aberration. They are shown as a function of field position in Fig. \ref{fig:zernikes}, with an average rms error of 7 nm rms for defocus, 13 nm for astigmatism and 0 nm for spherical in the inner region of the field of view until 0.5$^\circ$. Defocus reaches 45 nm rms at 1.0$^\circ$, while the astigmatism reaches 20 nm and spherical reaches 18 at the same field distance. The results reflect the overall wavefront error distribution from Fig. \ref{fig:contour-total}: the graphs are flat out to a distance of 0.4$^\circ$, beyond which they continuously rise beyond 100 nm rms after 1.0$^\circ$. The linear nature of the aberration modes confirm the linear dependence of the Zernike terms for a given field of observation as a function of the misalignment of L2, as it was demonstrated in paper II. While the big contributions of defocus and astigmatism repeat the results from paper II and paper III (where the wavefront analysis was done without the segmented mirror in the system), the spherical aberration was not expected to be this dominant. It is not clear at this point what causes it, especially since we would expect to see more significant coma in the off-axis PSF positions instead.
   
\subsection{Comparison to previous results without the segmented deformable mirror}

In paper II, figures 8, 9 and 10 show the design-predicted Zernike coefficients as a function of the field angle. Comparing our results to those, we can confirm the field-dependent rise of defocus and astigmatism. While paper II does show an increasing amount of spherical aberration in the wide field, it is not as dominant as the astigmatism while our results show the two to contribute equivalently to the wavefront error. In addition, we do not see coma showing up in our analysis, while we would expect to see some the further outwards we move.

Paper III has shown similar results in its figures 2 (experiment) and 4 (simulation), although the simulations have shown only significant defocus and astigmatism appearing in the off-axis wavefronts, and no other modes. In that experiment, there was no coma detected except in one of the corner PSFs, consistent with our non-detection in this work. The authors hypothesized that it was introduced by a flawed behavior of the L2 motors, since it showed up in only one of the four corners. This makes us confident that our detections of defocus and astigmatism in the present paper are real; however we are not able to tell at this point why the spherical aberration is so strong.

These findings support the further development of JOST into a multipurpose testbed that provides the possibility to implement different wavefront sensing and control techniques. With further work in the upcoming months, we will be able to provide new wide-field evaluations through the implementation of new WFS\&C algorithms, like the JWST baseline Hybrid Diversity Algorithm (HDA), Geometric Phase Retrieval (GPR), Optimized Phase Retrieval Algorithm (OPERA) and Estimation of Large Amplitude Subaperture Tip-tilt by Image Correlation (ELASTIC).



\section{SUMMARY AND CONCLUSION}
\label{sec:CONCLUSION}

JOST is a hardware simulator designed to test and validate wavefront sensing and control algorithms on segmented apertures like that of the JWST. In these proceedings, we presented the hardware and software updates performed on JOST since late 2017 and showed the characterization of the JOST wide-field with LAPD wavefront sensing.

One of the main hardware updates is the implementation of a new CMOS camera that allows for a faster image acquisition, but it reduced the overall field of view to about a quarter of the initial area. The second major hardware update is the new calibration flat map of the segmented deformable mirror, which was achieved by tweaking each individual segment in the pupil in piston, tip and tilt, until an overall surface error of 10 nm rms was achieved and integrated as the new baseline flat map on JOST.
On the software side, we migrated JOST to common Python tools in the Makidon Optics Laboratory. We put all code on version control, translated control and analysis scripts from Mathematica and IDL to Python, and make use of the HiCAT Python package for hardware control. The updated testbed is now a modular setup for wavefront sensing and control experiments.

We presented wavefront sensing results with the currently implemented LAPD wavefront sensing algorithm, going one step further in the validation of the wide-field optics since the implementation of the DM. We showed the overall wavefront error on field points ranging from -0.3$^\circ$ to 1.0$^\circ$ and presented the contributions from different Zernike modes. The total wavefront error rms rises from a nominal 40 nm until a field point of 0.4$^\circ$, where it starts to increase continuously, to over 100 nm at 1.0$^\circ$. The three main Zernike contributors to the wavefront error are Z4, Z5 and Z11, which are defocus, astigmatism and spherical aberration. While the defocus and astigmatism were expected from the analysis in paper II\cite{choquet2014} and III\cite{egron2016}, the spherical aberration has not occurred this strongly before and our upcoming work will investigate the cause of it. We expect further results of the JOST wide-field characterization by incorporating the Zemax interface within the JOST code to run simulations of the setup.

Now that the infrastructure is in place, with the L2 alignment and the Iris AO alignment done individually and a wide field evaluation and validation of the wavefront sensing,  we can proceed to investigate wide-field control solutions and comparison of multiple phase retrieval techniques, namely the Hybrid Diversity Algorithm (HDA), Geometric Phase Retrieval (GPR), Optimized Phase Retrieval Algorithm (OPERA) and Estimation of Large Amplitude Subaperture Tip-tilt by Image Correlation (ELASTIC).

\acknowledgments 
This work is supported by the JWST Telescope Scientist Investigation NASA Grant NNX07AR82G (PI: C. Matt Mountain).  This works also builds in part on work supported by NASA Grants NNX12AG05G and NNX14AD33G issued through the Astrophysics Research and Analysis (APRA) program for the development of the HiCAT testbed (PI: R. Soummer). This research made use of POPPY, an open-source optical propagation Python package originally developed for the James Webb Space Telescope project \cite{perrin2012}. This research made use of Astropy, a community-developed core Python package for Astronomy\cite{astropy2018}.

\bibliography{bibliotheque}
\bibliographystyle{spiebib}

\end{document}